\documentclass[journal,a4paper]{IEEEtran}

\usepackage[a4paper, margin=18mm]{geometry}

\usepackage{xfrac}
\usepackage{amsmath,amsfonts}
\usepackage[T1]{fontenc}
\usepackage{newtxtext,newtxmath}
\usepackage{booktabs}

\usepackage{graphicx}
\graphicspath{{images}}

\usepackage{listings}

\DeclareFixedFont{\ttb}{T1}{txtt}{bx}{n}{8} 
\DeclareFixedFont{\ttm}{T1}{txtt}{m}{n}{8}  

\usepackage{xcolor}
\definecolor{deepblue}{rgb}{0.17,0.14,0.53}
\definecolor{deepred}{rgb}{0.43,0.09,0.07}
\definecolor{deepgreen}{rgb}{0,0.5,0}

\newcommand\pythonstyle{\lstset{
language=Python,
basicstyle=\ttm,
morekeywords={self,None},              
keywordstyle=\ttb,
emph={mglyph,mg,Canvas,canvas,x},          
emphstyle=\ttb\color{deepred},    
stringstyle=\color{deepblue},
showstringspaces=false,
literate={-}{{\ttb-}}1 {=}{{\ttb=}}1, 
xleftmargin=1mm
}}

\lstnewenvironment{python}[1][]
{
\pythonstyle
\lstset{#1}
}
{}

\newcommand\pythoninline[1]{{\pythonstyle\lstinline!#1!}}

\definecolor{punct}{rgb}{0.12,0.2,0.1}
\definecolor{delim}{rgb}{0.15,0.15,0.15}

\lstdefinelanguage{json}{
    basicstyle=\ttm,
    numberstyle=\scriptsize,
    stringstyle=\color{eclipseKeywords}, 
    stepnumber=1,
    string=[s]{"}{"},
    stringstyle=\color{deepblue},
    showstringspaces=false,
    literate=
     *{0}{{{\color{deepred}0}}}{1}
      {1}{{{\color{deepred}1}}}{1}
      {2}{{{\color{deepred}2}}}{1}
      {3}{{{\color{deepred}3}}}{1}
      {4}{{{\color{deepred}4}}}{1}
      {5}{{{\color{deepred}5}}}{1}
      {6}{{{\color{deepred}6}}}{1}
      {7}{{{\color{deepred}7}}}{1}
      {8}{{{\color{deepred}8}}}{1}
      {9}{{{\color{deepred}9}}}{1}
      {.}{{{\color{deepred}{.}}}}{1}
      {...}{{{\color{punct}{\ \ \ \ $\cdots$}}}}{1}
      {:}{{{\color{punct}{:}}}}{1}
      {,}{{{\color{punct}{,}}}}{1}
      {\{}{{{\color{delim}{\{}}}}{1}
      {\}}{{{\color{delim}{\}}}}}{1}
      {[}{{{\color{delim}{[}}}}{1}
      {]}{{{\color{delim}{]}}}}{1},
}

\lstnewenvironment{json}[1][]
{
\pythonstyle
\lstset{#1}
}
{}

\usepackage[most]{tcolorbox}
\tcbset{
    frame code={},
    center title,
    left=0pt,
    right=0pt,
    top=0pt,
    bottom=0pt,
    colback=gray!10,
    colframe=white,
    width=\dimexpr\linewidth\relax,
    enlarge left by=0mm,
    boxsep=7pt,
    arc=4pt,outer arc=10pt,
    }

\usepackage{xcolor}

\usepackage[breaklinks,colorlinks]{hyperref}
\hypersetup{
	breaklinks=true,
	pdfpagemode={UseOutlines}, 
	plainpages=false, 
	pdftitle={The Malleable Glyph (Challenge)}, 
	pdfauthor={Herout et al.}, 
	pdfkeywords={}, 
	linkcolor=hrcolor-links, 
	citecolor=hrcolor-cite, 
	urlcolor=hrcolor-urls,
	colorlinks=true
}

\usepackage[noabbrev,capitalise,nameinlink,]{cleveref}

\usepackage{color}
\definecolor{hrcolor-links}{HTML}{af2831}
\definecolor{hrcolor-urls}{HTML}{092EAB}
\definecolor{hrcolor-cite}{HTML}{2F8F00}
\begin{document}

\title{The Malleable Glyph (Challenge)}

\author{Adam Herout, Vojtěch Bartl, Martin Gaens, Oskar Tvrďoch%
\\\emph{Graph@FIT, Brno University of Technology}%
\\[4pt]e-mail:~\href{mailto:herout@vutbr.cz}{herout@vutbr.cz}}

\markboth{Herout \MakeLowercase{\textit{et al.}}: The Malleable Glyph Challenge}%
{Herout \MakeLowercase{\textit{et al.}}: The Malleable Glyph Challenge}

\maketitle

\begin{abstract}
Malleable Glyph is a new visualization problem and a public challenge.  It originated from UX research (namely from research on card sorting UX), but its applications can be diverse (UI, gaming, information presentation, maps, and others).  Its essence is: carrying as much information in a defined planar and static area as possible.  The information should allow human observers to evaluate a pair of glyphs into three possible sortings: the first is ``greater'', or the second is ``greater'', or both are equal.  The glyphs should adhere to the \emph{illiteracy rule}, in other words, the observer should ask themselves the question ``how much?'' rather than ``how many?''.   This article motivates the technique, explains its details, and presents the public challenge, including the evaluation protocol.  

The article aims to call for ideas from other visualization and graphics researchers and practitioners and to invite everyone to participate in the challenge and, by doing so, move scientific knowledge forward.
\end{abstract}


\section{Introduction --- What is Malleable Glyph?}

\begin{figure}[b]
    \centering
    \includegraphics[width=0.8\linewidth]{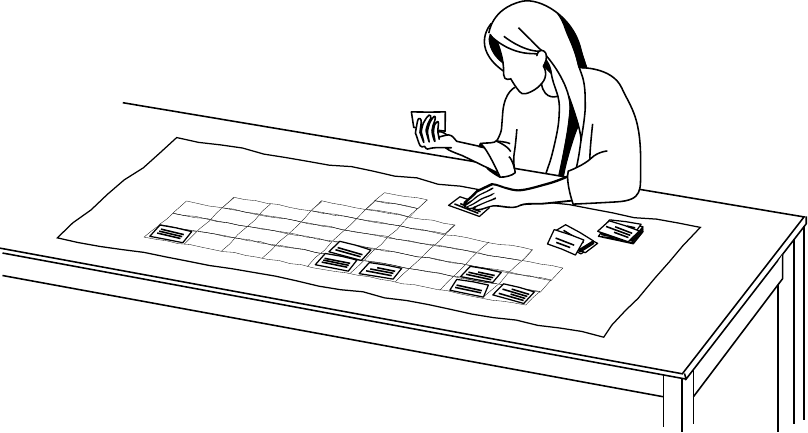}
    \caption{Illustration of the Q-methodology: subject sorts a deck of cards with subjective utterances into a predefined area typically shaped according to normal distribution. (illustration credit: Adriana Buchmei)}
    \label{fig:q-sorting}
\end{figure}

\IEEEPARstart{T}{he} origin of this research lies in the development of a modern user interface for the Q-methodology~\cite{Brown1993-Q-primer, Herrington2011-Q-overview}, which is a standard method for studying subjective opinions. Q-sorting was used long before the advent of computers and graphical interfaces, as shown in \cref{fig:q-sorting}. The essence of Q-sorting is to sort a deck of cards (tens to somewhat over a hundred) on a given scale, for example, from \emph{``does not describe me at all''} to \emph{``describes me exactly''}, or from \emph{``completely disagree''} to \emph{``fully agree''}. The cards contain phrases or short utterances to which the research participant is asked to take a position and rate the cards accordingly. The responses of the participants are then analyzed using factor analysis. 

In developing a modern user interface and seeking an optimal user experience, it was appropriate to remove the subjectivity that Q-sorting usually explores -- that is, to remove subjective statements from the sorted cards and replace them with objective content. This leaves only ``the second subjectivity'' -- the usability of the user interface.  
A good candidate for a deck of cards without subjectivity seemed to be cards with numerals written in text.  Working with these cards in sorting is at most similar to working with normal cards with subjective statements.  The participant reads the short text on the cards and is able to make a comparison between two cards as well as determine which card will be ranked ``very low'', which ``very high'' and which somewhere in the middle.  The trap, as our testing showed, turns out to be that human beings rank the numbers the way they learned to in elementary school.  First by the digit in the highest order, and then, within a group of cards with the same highest order value, by the next digit, and so on.  The way of working with cards that contain digits turned out to be completely different from working with cards that contain values (statements) that have no orders or digits.

\begin{figure}[h]
    \centering
    \includegraphics[width=0.8\linewidth]{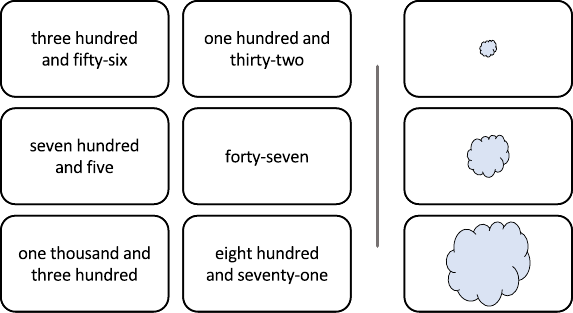}
    \caption{Card decks for UX testing of the card sorting user interfaces: \textbf{left:} cards with textual cardinals, \textbf{right:} cards with a scaled symbol.}
    \label{fig:q-sorting-motivation}
\end{figure}

The limitations of cards with numbers written in words led us to use a scaled geometric shape in the center of the card, as shown in \cref{fig:q-sorting-motivation}.  This solved the problem of numbers and ordering by discrete orders.  However, it turned out that there are not many sizes that the human eye can distinguish on a card of about $2\times1\,\mathrm{inch}$, and when testing with larger numbers of cards (dozens to hundreds), there are no longer enough sufficiently distinguishable sizes of shapes on the cards. 

\begin{figure}[t]
    \centering
    \includegraphics[width=\linewidth]{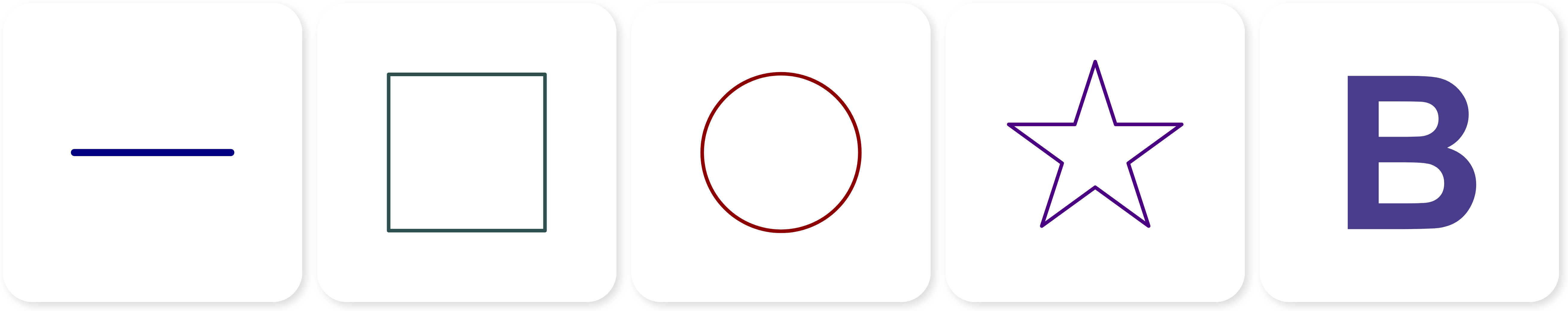}\hfill
    \caption{Preliminary testing showed that different scaled shapes (line, square, circle, star, character shape) have varying degrees to which a human subject recognizes their sizes.}
    \label{fig:just-scaled-glyphs}
\end{figure}

We conducted preliminary testing that, while not meeting the standards to be taken seriously statistically, showed that different geometric shapes allow human subjects to recognize their size to varying degrees.  The most basic shapes tested are shown in \cref{fig:just-scaled-glyphs}. Unsurprisingly, two-dimensional shapes offer more clues for size comparisons than a mere line segment. A circle offers better resolution capability than a square, proposing the hypothesis that a curved shape offers the eye more to lean on than a line segment. A star is comparable in resolution to a circle -- it is only composed of lines, but it contains more of them and contains different angles. Of the simple test shapes shown in the figure, the scaled letter B seemed to behave best, as it was chosen to contain both segments and curves with different curvatures.

This experimentation led to the \textbf{basic question of malleable glyphs}: How best to graphically use a given area so that, at first glance, a human subject can compare two glyphs with each other or, for one or more glyphs, determine where approximately they lie on a scale from smallest possible to largest?  For simplicity, we set the area to be a square of $1\times1\,\mathrm{inch}$, though other sizes may and will be studied.  Malleable glyphs are inherently scalar because they are intended to allow unambiguous pair ordering and placement on a one-dimensional scale. We introduce the convention that a malleable glyph is controlled by a real number parameter $x \in [0.0, 100.0]$ (numbers between $0$ and $1$ would work just as well but tend to be less intuitive for humans).

\begin{figure}[b]
    \centering
    \includegraphics[width=\linewidth]{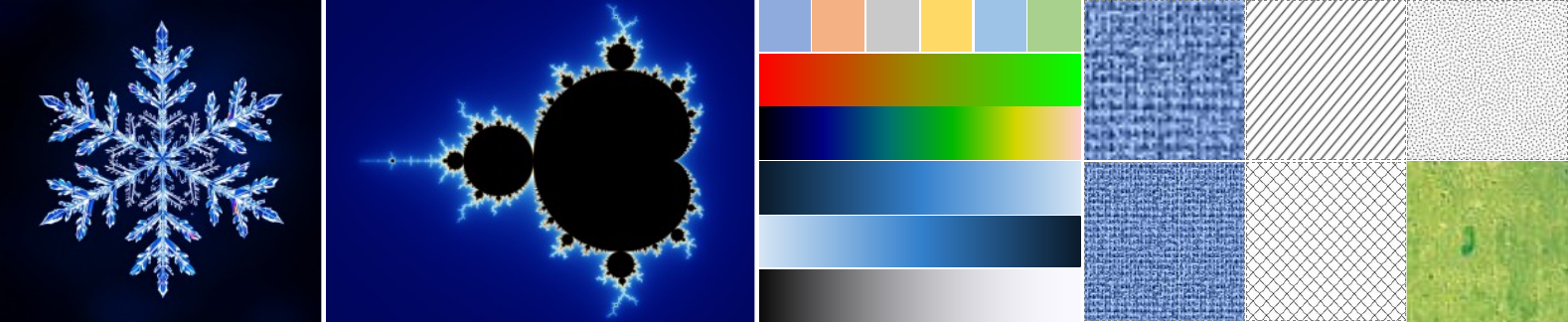}
    \caption{Malleable Glyphs do not have to be limited to using simple geometric shapes and scaling them. They can use any visual elements: raster graphics, fractal geometry, colors and color scales, textures, clouds of simple discrete elements, as well as anything else. (image credit: \href{https://stockcake.com/i/crystalline-winter-magic_1511734_1170668}{snowflake}, \href{https://commons.wikimedia.org/wiki/File:Mandel_zoom_00_mandelbrot_set.jpg}{mandelbrot})}
    \label{fig:various-graphics}
\end{figure}

We originally wanted to call the glyphs we studied \emph{scalable} glyphs, but that would imply that they should only be created by scaling vector geometric shapes (as they even originally were).  However, such a suggestion would be misleading: glyphs can use any graphical and visual elements, provided they fit within the specified dimension and conform to other rules (in particular, the Illiteracy Rule described in \cref{sec:illiteracy-rule}).  The originally intended use of glyphs is for printing on physical cards, so we do not allow the use of animations and temporally dynamic graphical elements.  However, studying exactly such animated \emph{videoglyphs}, as a modification or special category of malleable glyphs also makes sense. Of the static graphic and visualization elements, anything can be used; see \cref{fig:various-graphics}.  Of course, it is possible to use color -- both to achieve aesthetic quality and as a means of expression of the glyph's magnitude, perhaps through color gradients.  The use of fractal geometry can be very promising, making it possible to vary the \emph{complexity}, not just size, of geometric shapes, either continuously or in a high number of discrete steps.  Malleable glyphs can also use a variety of textures and clouds or clusters of points or other simple elements~--~or anything else.

\subsection{Malleable Glyphs are all around us}

Since way back, people have used graphical elements to represent some numerical, quantitative quality -- typically in illustrations and charts. Weber and, subsequently, Fechner~\cite{Fechner1860} studied the perception of the size of geometric shapes as one of the modalities of the logarithmic law of perception.  The advent of computer graphics made it possible to generate shapes automatically and quickly according to precise numerical specifications, and researchers began to study the perception and usability of individual (then simple, mostly line-based) glyphs~\cite{Mackinlay1986}.  A more recent state-of-the-art report by Borgo et~al.~\cite{Borgo2013} gives a fairly detailed overview of research in this area.

More recent work in finding glyphs suitable for vi\-su\-a\-li\-zing scalar information and for visualizing several numerical quantities at once typically focuses on a narrow class of glyphs and relates them to a specific task.  For example, Opach et~al.~\cite{Opach2018} search for the optimal glyph for a given task of visualizing multiple quantities at a single point, still using strongly geometric (circle, polyline, \dots) glyphs.  Fuchs et~al.~\cite{Fuchs2013} construct relatively complex glyphs from filled squares for visualizing multidimensional information and perform a comparative study of their perception by a human subject.  Also, in recent research, Raidvee et al.~\cite{Raidvee2020} address the much-studied problem of perceiving the proportions of basic shapes (variants of a circle), their diameter, area, merging two values into one and ``correct'' visualization of such merging, as well as estimating the mean value of two or more glyphs.  This work also provides a good review of the previous literature on this topic and contains a suitable methodology for carrying out similar subjective studies.

Glyphs seem to have received special attention in the context of visualization using \emph{scatter plots}. Glyphs in this context are used both as value carriers (typically through their size) and without a scalar value, merely serving to distinguish individual markers in a scatter plot.  Li et~al.~\cite{Li2010} specifically studied the use of glyphs with a scalar parameter (namely, a circle with a variable radius) as a visualization and differentiation tool in scatter plots.  Burlinson et~al.~\cite{Burlinson2018} then studied the use of shapes other than the conventional circle.  Their study contrasts open vs. closed shapes, where a representative of open shapes might be a cross or a star composed of line segments, and simple closed shapes are the circle, square, and triangle.  This study makes use of the Flanker paradigm~\cite{Forster2007} and presents interesting findings on the applicability of glyphs (which do not serve as carriers of scalar value but only of mutual resolution) in scenarios with high cognitive load and with relatively complex tasks through a large number of perceived features.  Similarly, Gleicher et~al.~\cite{Gleicher2013} use glyphs without a scalar value, only serving for mutual differentiation, and compare their utility with the use of color, still in the context of scatter plots.

An important use of glyphs carrying scalar (or multi-scalar) information and allowing intuitive relative comparisons with each other lies in the field of maps~\cite{MacEachren1995}.  Müller et~al.~\cite{Müller2023} have relatively recently proposed a technique called ``square-glyphs'' used to visualize multi-dimensional information (about 4 dimensions) in the context of maps.  A square-glyph is composed of several adjacent squares of different colors.  The conclusion of the study is that visualization using this tool is effective; it is appropriate to add that it is probably not optimal, and the appearance of the glyphs used encourages the search for truly aes\-the\-ti\-cally beautiful visualization tools that are also effective in communicating scalar and/or multi-scalar information.  Even more complex glyphs for visualizing information in maps have been proposed by Scheepens et~al.~\cite{Scheepens2014}, who focused on the visualization of moving objects and their multi-scalar properties (specifically the movement of ships in and around a harbor).  The authors propose new glyphs based on pie charts but try to minimize their size to avoid overlapping at exposed locations in the map while preserving the information conveyed.

Interesting work by Ying et~al., GlyphCreator~\cite{Ying2021} uses machine learning to generate glyphs for visualizing scalar or multi-scalar information based on an example.  Their glyphs focus on circular glyphs (mostly pie charts and their modifications) and advanced aesthetics around them.  Also worth mentioning is the relatively recent work of Zhao et~al.~\cite{Zhao2022}, who design novel 3D glyphs; the work focuses on visualizing atomic spin, so the authors are interested in visualizing two values that are perfectly separable and yet visualized by a single, intuitively readable glyph.

As glyphs in our sense, many different user interface elements can be considered, which in some cases have been studied in detail scientifically, but in many cases have not and were just casually designed and seem to be working.  A computer user interface contains many scroll bars, progress bars, and scalar value indicators (for example, for volume or brightness and many others).  Similarly, the interfaces of cars, planes, and machines contain indicators of various quantities, which are supposed to be readable without decoding digits and, in addition to reading the absolute value, are supposed to allow peripheral perception of changes in quantities and notice, especially sudden and rapid changes.  A great deal of creativity in the design of different glyphs (in the spirit of our definition) has been devoted to the design of computer games of different genres to visualize the properties of different entities in the game.

Our goal is to offer a space where the diverse experiences of designers and researchers studying related visualization problems meet.  We want to stimulate artistic and design creativity and create a sufficiently diverse zoo of malleable glyphs.  In doing so, we still want to maintain the ability to evaluate and compare glyphs objectively.  The attempt to reconcile these two requirements leads to the proposed definition of the malleable glyph; the definition is to be minimalist, clear, and restrictive in fundamental principles to form a clear ``playground'', but deliberately as loose as possible in detail and in the graphic design to leave maximum room for creativity.

\section{Evaluation of Malleable Glyphs}
\label{sec:glyph-evaluation}

\begin{figure}
    \centering
    \includegraphics[width=0.45\linewidth]{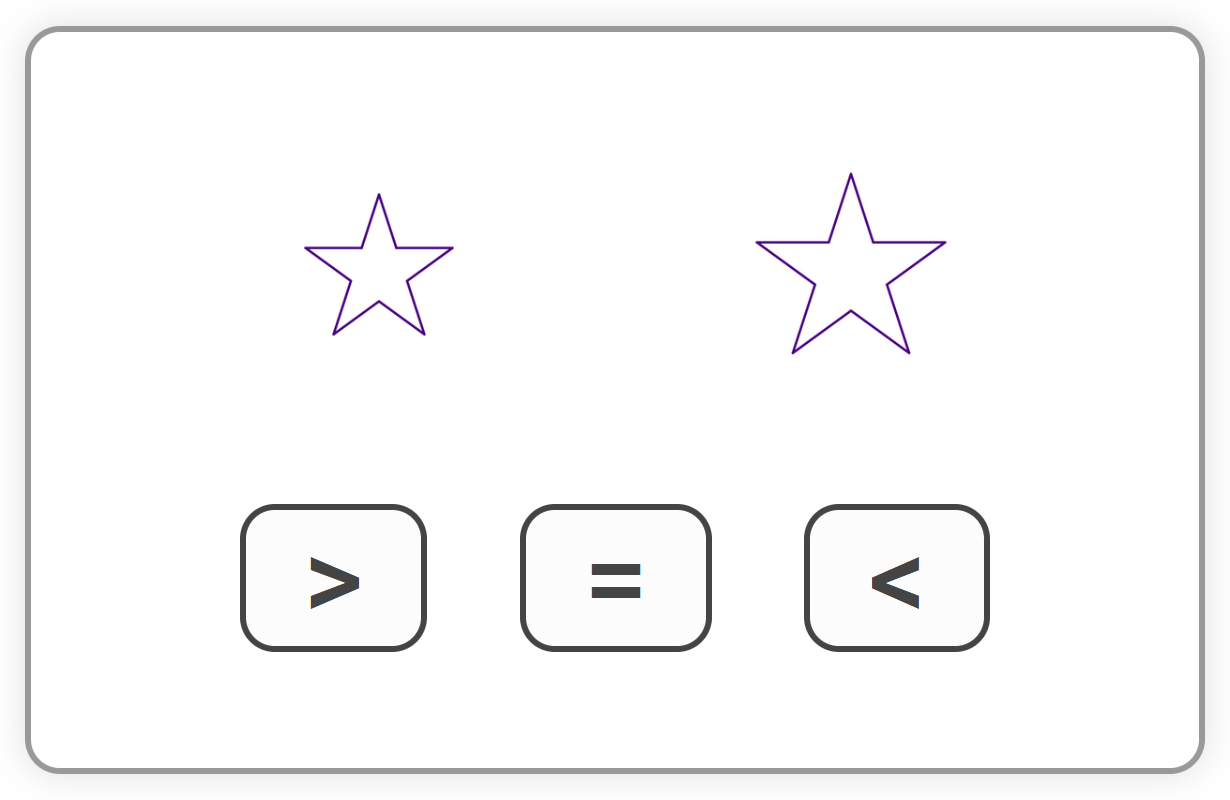}
    \caption{Evaluation UI central element: the human subject is given two glyphs of the same kind with different or equal parameters $x$ and makes a decision.}
    \label{fig:evaluation-ui}
\end{figure}

The main measurable quality of a malleable glyph is its resolution, i.e. how small a difference in the parameter $x$ the observer is able to detect, or how many different ``shades'' of the same glyph the observer is able to distinguish.  To measure this property, we propose to use pairwise comparisons of pairs of the same glyph with different (or the same) parameter $x$, using the simple user interface shown in \cref{fig:evaluation-ui}. 

During one experimental session, the observer can be evaluating one or multiple different glyph designs which are presented to them mixed in a random order. For each of these glyph designs, the test application presents the observer with a pair of glyphs of sizes $x_1$, $x_2$ ($x_1$ on the left, $x_2$ on the right):
\begin{equation}
\begin{aligned}
    x_1 &= c \pm \sfrac{d}{2} \\
    x_2 &= c \mp \sfrac{d}{2} ,
\end{aligned}
\end{equation}
where $d$ is a suitably chosen \emph{distance} between the two glyphs under test and $c$ is a \emph{central} value generated randomly based on a uniform distribution 
\begin{equation}
    \label{eq:random-center-value}
    c \sim U(\sfrac{d}{2}, 100.0 - \sfrac{d}{2}). 
\end{equation}
Randomly, with probability $p_= = \sfrac{1}{3}$, glyphs with identical parameter $x$ are presented to the observer:
\begin{equation}
    x_1 = x_2 = c
\end{equation}
so that the answer choice ``='', i.e., the glyph on the left is equivalent to the glyph on the right, is not zero, but instead shares the same probability with the other two answers ``>'' and ``<''.

First, the observer is presented with a pair of glyphs with mutual distance $d_0$, which is chosen so that the observer is quite certain to detect the difference, typically $d_0 = 20.0$. In subsequent comparisons, the distance is gradually reduced to 
\begin{equation}
    d_t = \gamma^t d_0 ,
\end{equation}
where $t=0, 1, 2, \dots$ is a parameter of the individual test, and $\gamma$ determines the fineness of sampling of the spatial resolution, typically $\gamma=0.7$. If the observer judges the pair of glyphs correctly, a smaller distance is chosen for the next comparison
\begin{equation}
    \label{eq:t-increase}
    t \leftarrow t+1
\end{equation}
and if the observer's answer is incorrect, the experiment reverts back to ``safer'' distances, we propose 
\begin{equation}
    \label{eq:t-decrease}
    t \leftarrow \max(t-3, 0) .
\end{equation}
In fact, formula \labelcref{eq:t-increase} is only applied if $x_1 \neq x_2$; when two identical glyphs were presented to the observer, i.e. $x_1 = x_2$, and the answer was correct, the parameter $t$ will not change for the next comparison.

The proposed method of testing leads to the observer answering correctly about 80\,\% of the time because in this ratio, the observer is given a pair of glyphs to judge (the ratio is determined by formulas \labelcref{eq:t-increase} and \labelcref{eq:t-decrease}).  The just noticeable difference (i.e., the distance $d$ for which just half of the time the observer can make a correct decision) corresponds to a ratio of \sfrac{2}{3} correct decisions for a given $t$; if the observer were making completely random, completely uninformed decisions, the probability of a correct decision would be \sfrac{1}{3}.

All comparisons of glyphs in the subjective testing are recorded in database $\mathcal{R}$.  For each glyph $G$ and for each tested distance $d_T$ determined by the parameter $T=0, 1, 2, \dots$, the accuracy of the answers is calculated as
\begin{equation}
    \label{eq:accuracy}
    A(G, T) = \frac{|\mathcal{R}(G, t=T, \checkmark)|} {|\mathcal{R}(G, t=T)|}
\end{equation}
where $\mathcal{R}(G, t=T, \checkmark)$ is the subset of all responses $\mathcal{R}$ that pertain to glyph $G$, the distance $d_t$ was determined by parameter $t$ exactly equal to $T$ and the response supplied by the observer was correct; $\mathcal{R}(G, t=T)$ is the subset of all responses (both correct and incorrect) pertaining to glyph $G$ where the parameter $t$ is exactly equal to $T$.

\begin{figure}
    \centering
    \includegraphics[width=\linewidth]{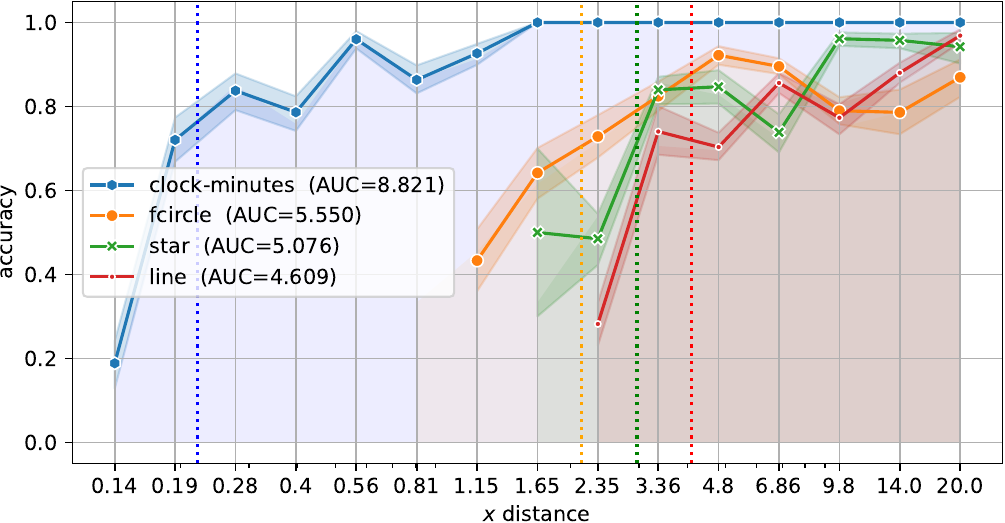}
    \caption{Plot of accuracies according to \cref{eq:accuracy} for four different glyphs.  It is clear that the \emph{clock with minute graduations} is much better than the others, which are mutually almost on par.  The line plot contains plots of confidence intervals computed by bootstrapping, which are very narrow.  Vertical dotted lines correspond to the \emph{resolution} of the glyphs or the \emph{just noticeable distance} for each glyph design.}
    \label{fig:accuracy-plot}
\end{figure}

An example of the evaluation of four different glyphs (\emph{line}, \emph{star}, \emph{circle}, and \emph{clock with minute graduations} as shown in \cref{fig:one-handed-clock}, bottom) based on \cref{eq:accuracy} is shown in \cref{fig:accuracy-plot}.  The accuracies of $A(G, T)$ for increasing $T$ (i.e., when reading the plot from right to left) may not always be decreasing or non-increasing, but by the nature of things, near the \emph{just noticeable difference}, they begin to decrease and eventually fall to an accuracy of \sfrac{1}{3}, which corresponds to random, completely uninformed, responses.  \cref{fig:accuracy-plot} includes the confidence intervals made by bootstrapping; it is notable that the confidence intervals are very narrow even for a relatively small number of observer's responses (708 answers for all the four glyphs, about 20 minutes spent by a single observer).


As a single numerical characteristic of the glyph -- instead of reading the entire curve from \cref{fig:accuracy-plot} -- the area under each curve appears to be suitable.  Somewhat unusually, we propose to compute the area under the curve on a logarithmic scale, just as is the plot.  The formula for calculating the area under the curve is as follows:
\begin{equation}
  \begin{aligned}
    \mathcal{A}(G) = \frac{1}{\ln2}\Bigg( 
      &\sum_{t=0}^\infty \frac{1}{2}\Big(A(G,t) + A(G,t+1)\Big) \ln\frac{1}{\gamma} \\
      &+ \ln\frac{100.0}{d_0} \Bigg).
  \end{aligned}
\end{equation}

The sum in the formula adds together the areas of the individual trapezoids constructed at the measured points~$d_t$, whose width in the logarithmic scale is $\ln\frac{1}{\gamma}$. The sum is formulated from $t=0$ to infinity, although in practice, for some $t$, the set of records $\mathcal{R}$ will contain no records for the given glyph, and the summation of areas can then be terminated because the area under the curve is zero for all higher $t$. The term $\ln\frac{100.0}{d_0}$ corresponds to the area of the curve from $d_0$ to $100.0$ so that the computed area under the curve is always comparable even for varying $d_0$; note that from $d_0$ to $100.0$ we assume perfect success, i.e., $d_0$ is indeed chosen so that at this distance of the glyphs being compared and at higher distances, the observer can certainly distinguish the glyphs correctly. 

The term $\frac{1}{\ln2}$ normalizes the computed area under the curve so that halving the distance of the glyphs corresponds to a unit distance on the horizontal, logarithmic scale. Then,
\begin{equation}
    R(G) = 2^{\mathcal{A}(G)}    
\end{equation}
is the glyph \emph{resolution}, that is the expected number of different values of $x \in [0.0, 100.0]$ that an observer is able to distinguish at the level of just noticeable difference. Value
\begin{equation}
    D(G) = \frac{100.0}{R(G)}
\end{equation}
is then the assumed \emph{just noticeable distance} between glyphs of the same design $G$.

The evaluation proposed in this section is based on the UI from \cref{fig:evaluation-ui} and assumes a temporally unrestricted exposure towards the observer and the two glyphs being positioned near each other.  It is a matter for future work to implement other exposure scenarios -- time-limited exposure, exposure sequentially in time at the same or different locations, greater distance between the glyphs being compared, positioning the glyphs such that the observer has to carry out eye or full head movement, and other scenarios.

\subsection{Calibration of glyphs in relation to the Weber-Fechner law}

The Weber-Fechner law is based on Weber's observation that the intensity of the just noticeable difference is proportional to the intensity of the original stimulus
\begin{equation}
    \frac{\Delta S}{S} = k ,
\end{equation}
where $\Delta S$ is the just noticeable difference, $S$ is the intensity of the original stimulus, and $k$ is a constant empirically found for a given stimulus type and its perception.  Fechner~\cite{Fechner1860} modified this observation into the formulation that the intensity of the perceived stimulus is proportional to the logarithm of the original stimulus:
\begin{equation}
p = k \ln \frac{S}{S_0} .
\end{equation}

\begin{figure}
    \centering
    \includegraphics[width=\linewidth]{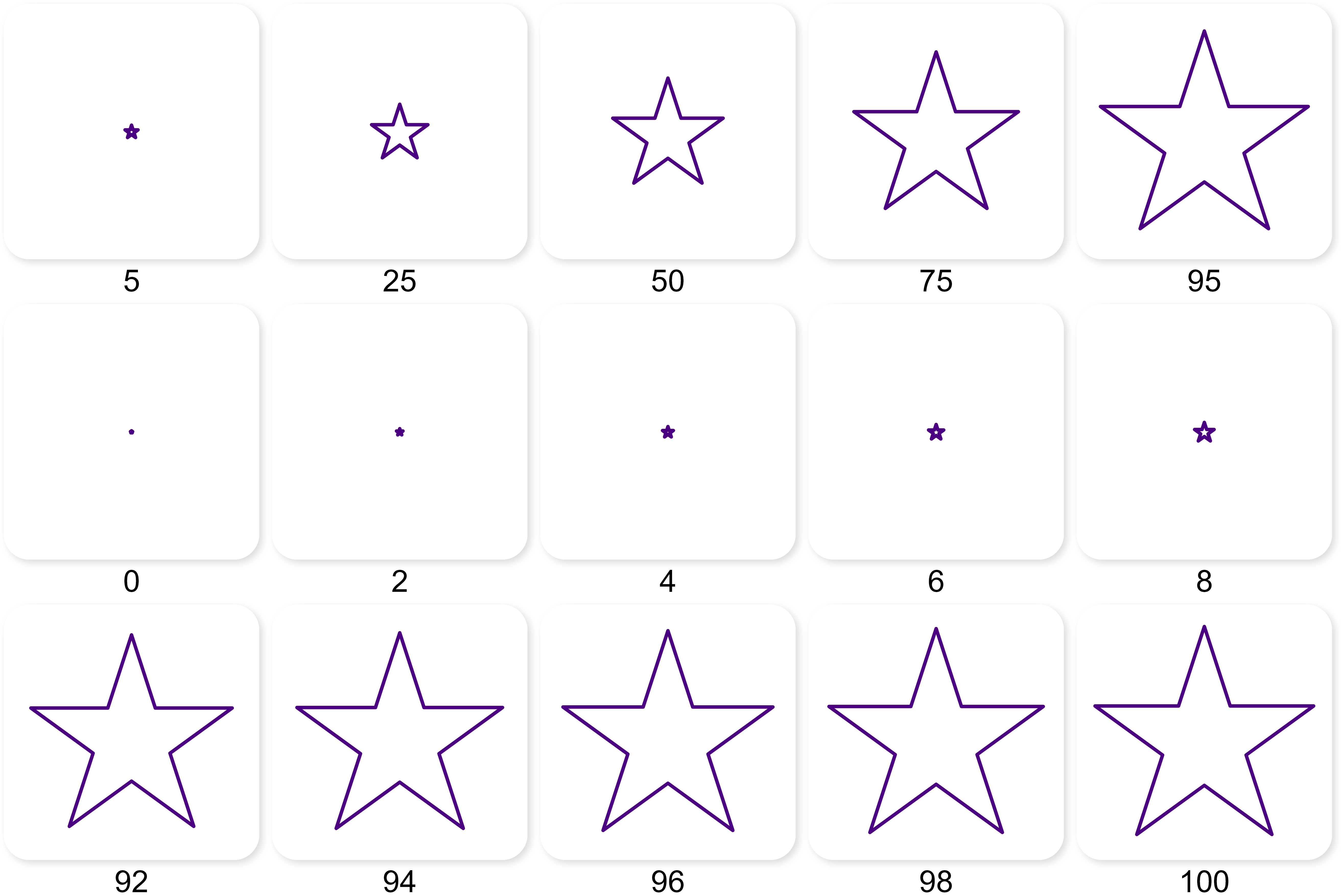}
    \caption{A simple glyph illustrating the Weber-Fechner law. The differences between glyphs with distance $d=|x_2-x_1|=2$ are much clearer for small values of $x$.}
    \label{fig:weber-fechner-star}
\end{figure}

\cref{fig:weber-fechner-star} shows that the Weber-Fechner law apparently applies to the perception of malleable glyphs as well -- for low values of $x$ in this particular case, it is easier to observe changes than for higher values.  Weber-Fechner's law applies to the perception of the size or scale of objects, or to the perception of the intensity of the brightness of a color, but, for example, it does not apply to the perception of the hue of a color.

The method of glyph evaluation proposed in this chapter presents glyphs to the observer randomly with a uniform distribution with respect to the entire space $x$ of glyphs, \cref{eq:random-center-value}. The resulting glyph characteristics, i.e., the curve $A(G,T)$, the area under it $\mathcal{A}(G)$ or the glyph resolution $R(G)$, are evaluated collectively as if the just noticeable difference between glyphs did not depend on the value of $x$.  It is up to the glyph designer to implement an analog of the gamma correction used in displays so that, if possible, the just noticeable difference is uniform over the entire range of $x$. By the fact that glyphs can combine all sorts of visual communication, and the subcomponents of a display can be \emph{phased over} different parts of the interval of $x$ values (see the next section and \cref{fig:composite-glyphs}), the Weber-Fechner law will not necessarily apply to the glyph as a whole, but rather to its subcomponents, and the calibration against this law will be specific to a particular glyph design.

\subsection{Evaluation of the aesthetic appeal of glyphs}

The preceding text describes the proposed numerical metric for evaluating the resolution of glyphs.  In addition to this numerical characterization, we aim to find malleable glyphs that are aesthetically beautiful.  We anticipate that it will also be interesting to study the trade-off between aesthetic appearance and glyph resolution, and ways to reconcile these two -- apparently somewhat conflicting -- requirements.

In line with the findings of Mantiuk et al. \cite{Mantiuk2012} and other works, we consider it appropriate to use the \emph{forced-choice pairwise comparison} method, using the existing methodology and software \cite{PerezOrtiz2017}.  The subject of glyph aesthetic appeal testing can be exposed in each pairwise comparison not just by two images of two different glyphs, but each of the glyphs in the pairwise comparison can be represented by a small number of realizations of the glyph at different $x$.  An accurate definition of the experimental methodology and the creation of a self-evaluation tool and a tool for evaluation within the challenge of assessing aesthetic appeal is one of the tasks for future work.

\section{The Illiteracy Rule}
\label{sec:illiteracy-rule}

\begin{figure}
    \centering
    \includegraphics[width=\linewidth]{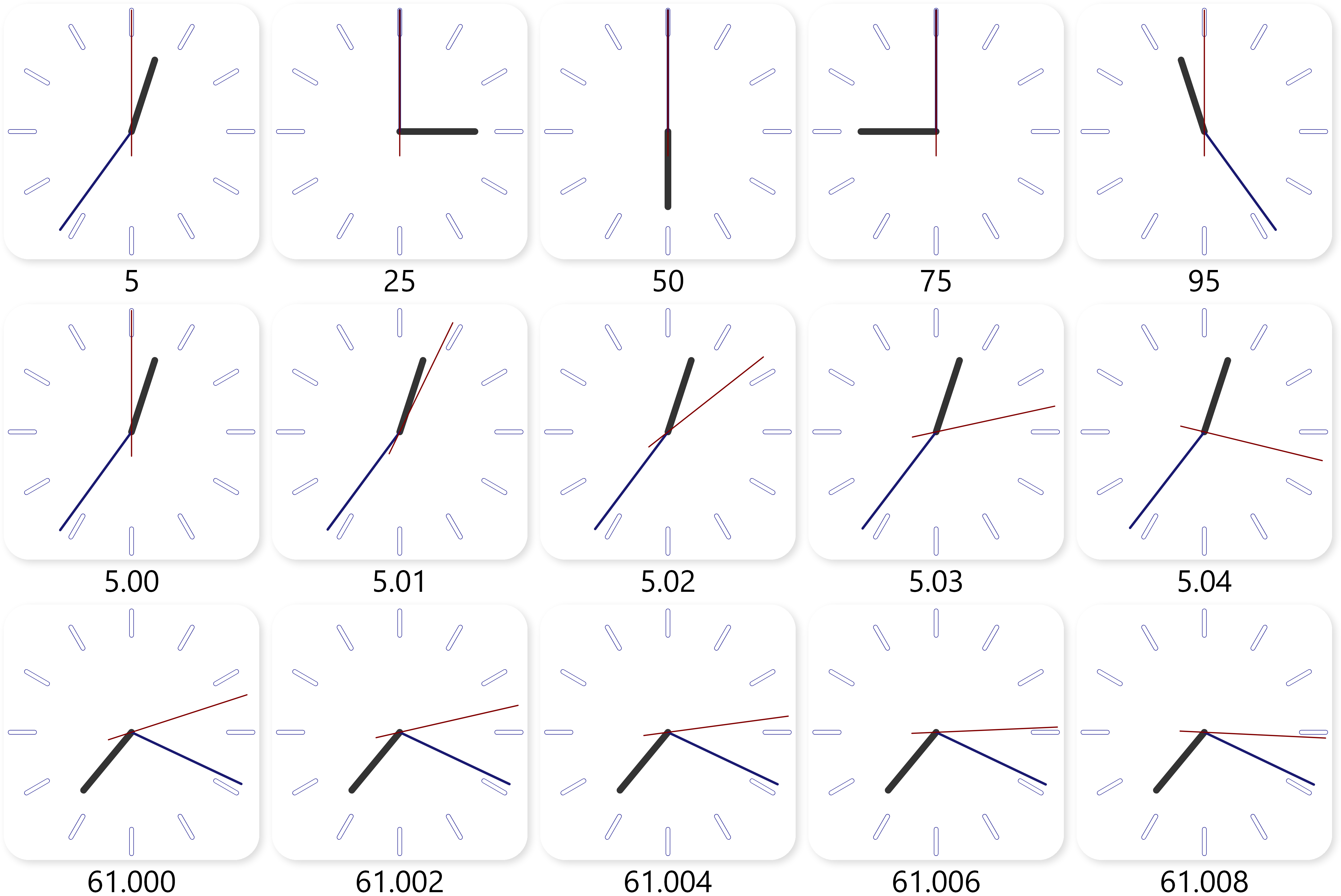}
    \caption{\textbf{Three-Handed Clock:} An example of a glyph that is extremely efficient (it is possible to clearly recognize differences of $0.002$ in the $x$ values), but violates the illiteracy rule.}
    \label{fig:three-handed-clock}
\end{figure}

\cref{fig:three-handed-clock} shows a three-handed clock that performs excellently as a malleable glyph: the bottom row shows that, based on the second hand, glyphs with a mutual distance $\delta x$ of $0.002$ and even $0.001$ can be distinguished reliably. A clock with four or five hands would, of course, work even better. The movement of the hands on the clock corresponds exactly to the meaning of the digits in the numerals, and the three-handed clock is thus a return to the numeral cards from \cref{fig:q-sorting-motivation} -- which have shown themselves as undesirable.

Therefore, we make it a rule for malleable glyphs that they must not use writing (both letters and numerals), nor must they use numerals or values arranged in individual \emph{orders} even without using writing -- as is the case with the hands of the three-hand clock. An apt name for this rule seems to be the \textbf{Illiteracy Rule} -- because the rule is intended to make the glyphs automatically intelligible to someone who is perfectly illiterate: such as an animal or a small child.

\begin{figure}
    \centering
    \includegraphics[width=\linewidth]{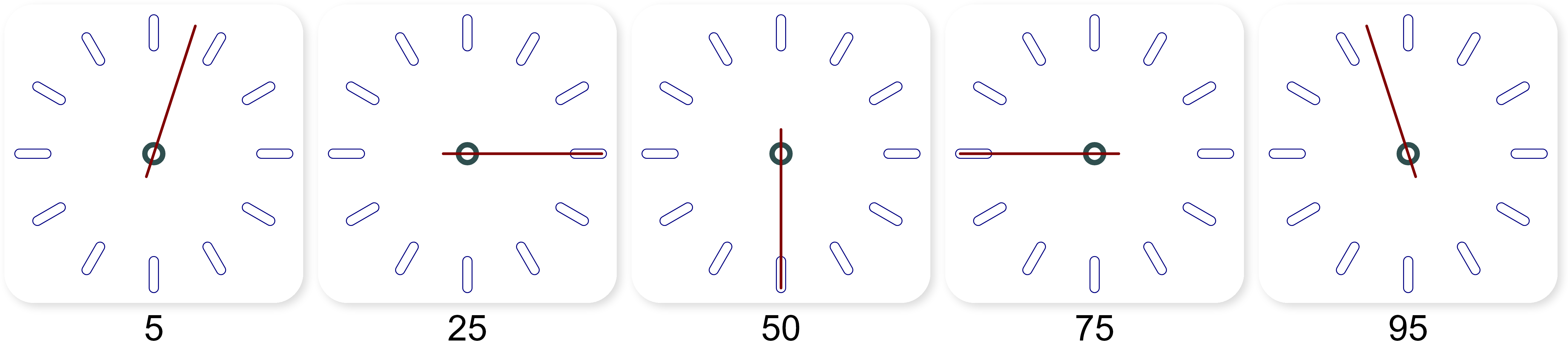}\\[1pt]
    \includegraphics[width=\linewidth]{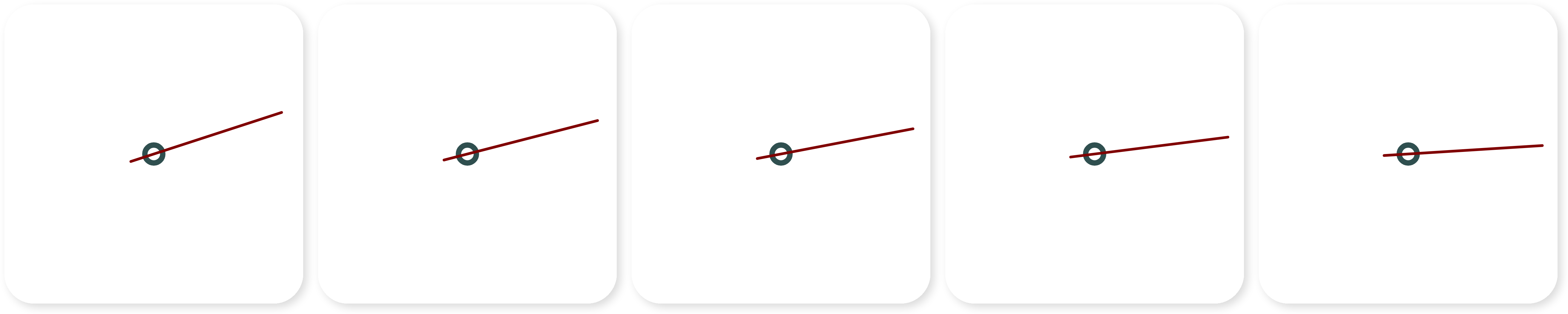}\\[1pt]
    \includegraphics[width=\linewidth]{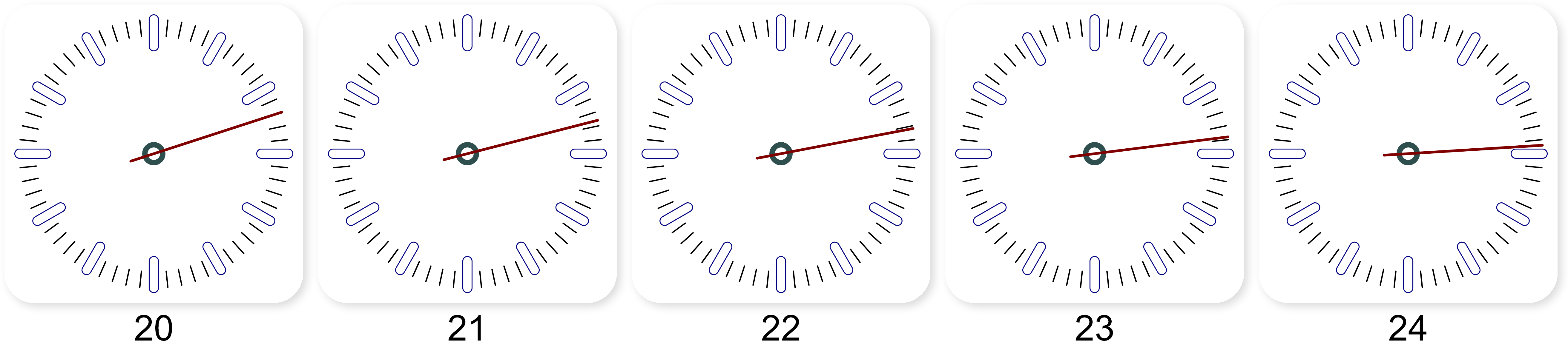}
    \caption{\textbf{One-Handed Clock:} This is an eligible malleable glyph because it does not use any digits or orders.  The use of graduations on the clock face does not violate the illiteracy rule and can help: different positions of the hand clock are clearer with the minute graduations than on an empty clock face.}
    \label{fig:one-handed-clock}
\end{figure}

The metaphor of the clock itself (as well as other metaphors) is not necessarily problematic. \cref{fig:one-handed-clock} shows a clock with a single hand, which may well make use of markers on the dial -- graduations -- to help the recognition of the position of the clock hand.

\begin{figure}
    \centering
    \includegraphics[width=\linewidth]{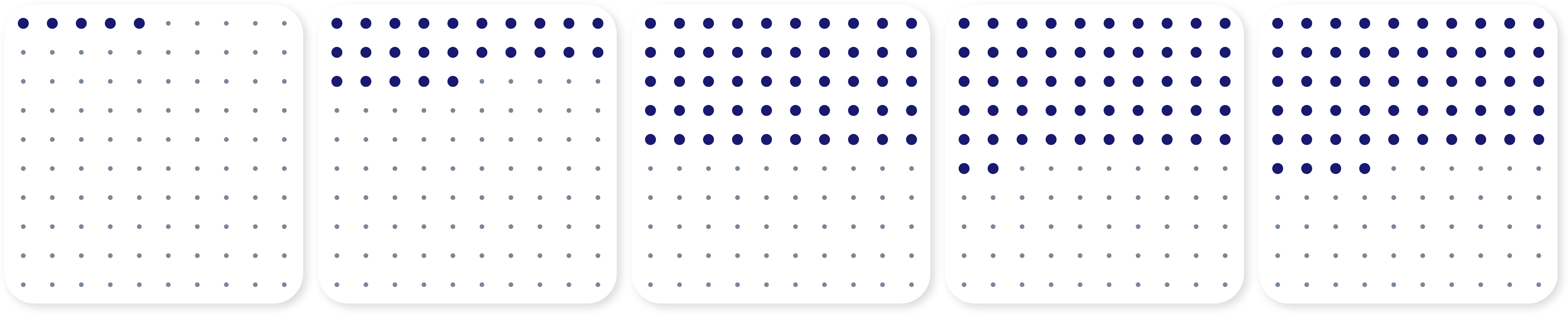}\\[1pt]
    \includegraphics[width=\linewidth]{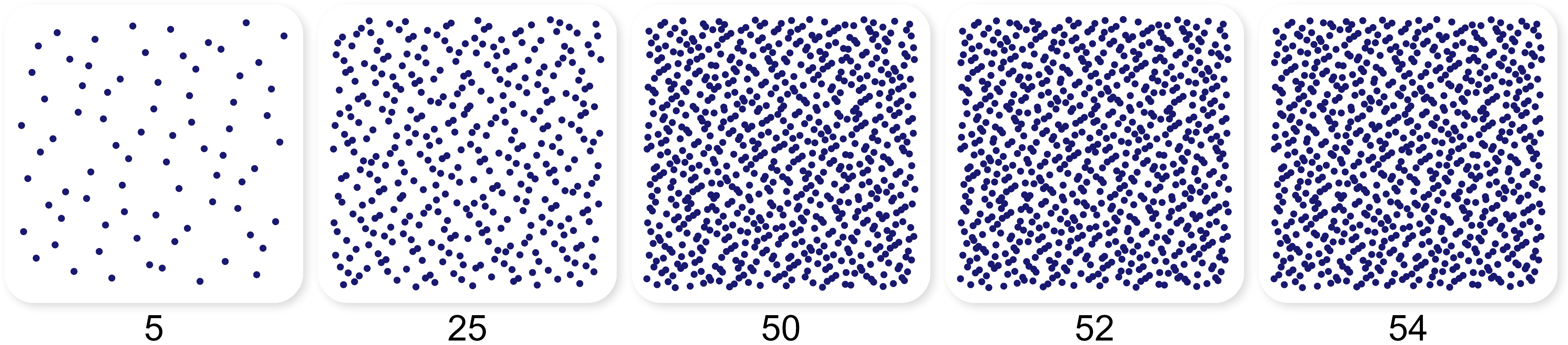}
    \caption{Invalid and valid use of discrete dots.  
      \textbf{top:} Dots distributed in regular rows form two \emph{digits} or \emph{orders}, violating the illiteracy rule. 
      \textbf{bottom:} Dots distributed in a pseudo-random way are fine.  The question for the observer should be ``how much?'' instead of ``how many?''}
    \label{fig:points}
\end{figure}

The principle of using digits that are not expressed by explicit writing can be illustrated by another simple example -- a dot matrix that the observer views as two digits: the number of complete rows is the higher order, the number of dots in the partial row is the lower order (\cref{fig:points}). When deciding on the ordering of two close glyphs, the observer must first be interested in the higher order, and only if there is agreement on that order, be interested in the lower order. The bottom row of the figure shows the use of the same discrete dots, which is legitimate because it does not violate the illiteracy rule. In the bottom row, the dots are distributed based on the pseudo-random Halton sequence~\cite{Halton1964}. This example suggests that an alternative formulation of the illiteracy rule might be that the observer must be confronted with the question ``how much?'' rather than the question ``how many?''

The idea of the Illiteracy Rule is related, though not identical, to the notion of \textbf{subitizing}~\cite{Kaufman1949}, which refers to the ability to count a small number of objects in a group quickly, effortlessly, reliably, and confidently.  Subitizing in humans works reliably with up to four objects; beyond this number, recognizing each additional object requires an additional investment of cognitive effort.  It is a matter for further research, once a sufficient variety of malleable glyphs have been collected and systematized, to apply the criteria and research methods used to describe subitizing -- in particular, exposure time constraints and confidence ratings.

\begin{figure}
    \centering
    \includegraphics[width=\linewidth]{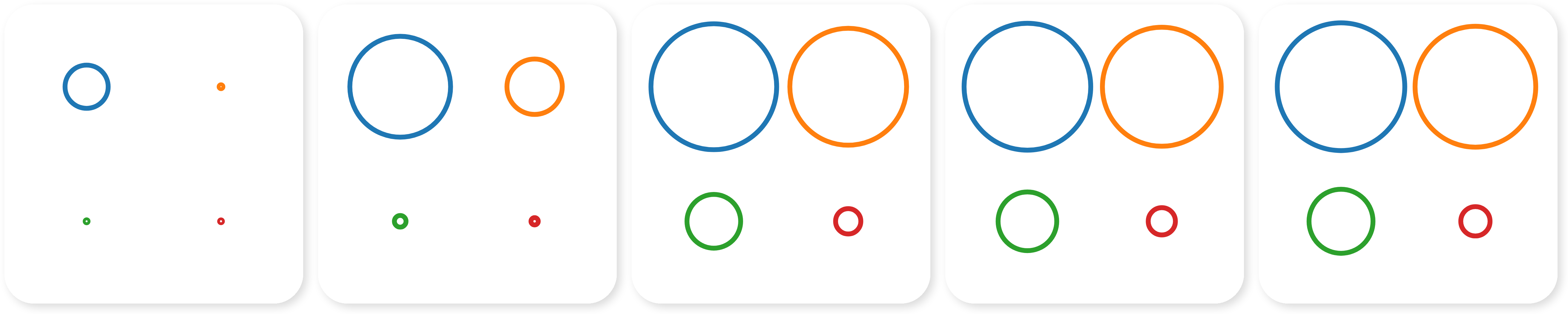}\\[1pt]
    \includegraphics[width=\linewidth]{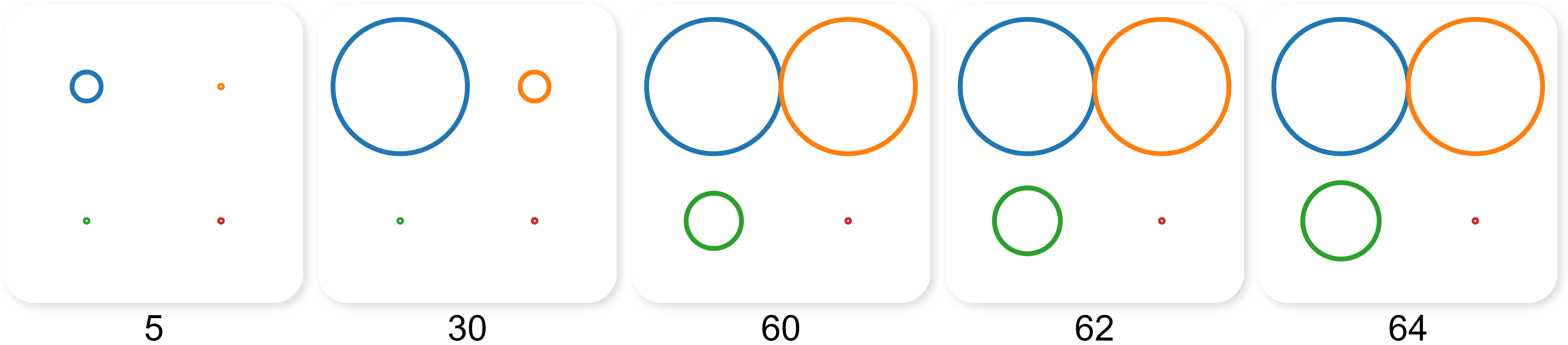}\\[6pt]
    \includegraphics[width=\linewidth]{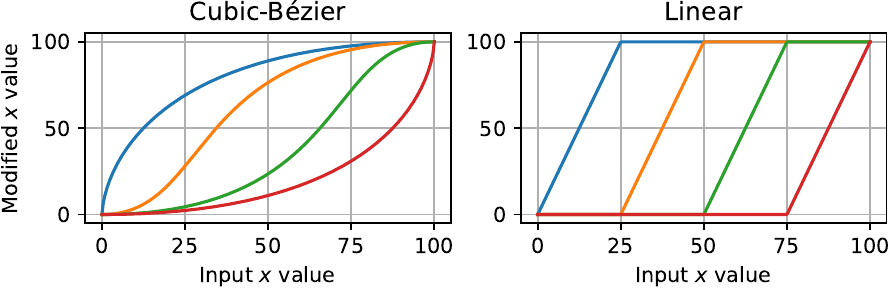}
    \caption{Simple composite glyphs composed of four circles. 
       \textbf{first line:} Circles are phased with cubic Bézier curves.
       \textbf{second line:} Circles are phased linearly.
       \textbf{plots:} Cubic Bézier and linear phasing of the circles -- as examples of phasing of the composite glyph parts.
       The evolvement of the cubic Bézier version seems to be more elegant and aesthetically pleasing, while the slope/gradient of the linear movement is equal across the domain of $x$ and overall higher.}
    \label{fig:composite-glyphs}
\end{figure}

Forbidding the use of numbers and orders may sound to some like all elements in a glyph must ``happen'' at once, linearly with respect to $x$. But this is not the case: a glyph can be composed of phenomena and features that have their own independent (or dependent) phasing. \cref{fig:composite-glyphs} shows a simple example where a glyph consists of four scaled circles, each with its own scaling rate. The figure shows two variants of the same composite glyph: the top row phases the circles based on cubic Bézier splines, while the bottom row phases the circles linearly. The benefit of phasing each part of the glyph is that the gradient of the change of a feature in the glyph (in this example, the scale of one circle) with respect to $x$ can be higher than in the case of a glyph that is not composite, and for which the feature change happens linearly with respect to $x$.

We are curious about composite glyphs that will make efficient use of the given bounded area and place individual composite features appropriately in space, and also combine different modalities (scale, fractal complexity, color, texture characteristics, etc.) with appropriate phasing. With respect to the Weber-Fechner law discussed earlier, it may be advantageous to stimulate the observer with multiple phased smaller elements to which they are relatively more sensitive, rather than with one large one whose relative changes will already be below the just noticeable difference threshold. We believe that the malleable glyphs offer a well-defined framework so that creative design work and parallel research on the interaction of partial influences can advance knowledge in the field of visualization and perception.

\section{Preliminary Categorization of Malleable~Glyphs}

\begin{figure}
    \centering
    \includegraphics[width=\linewidth]{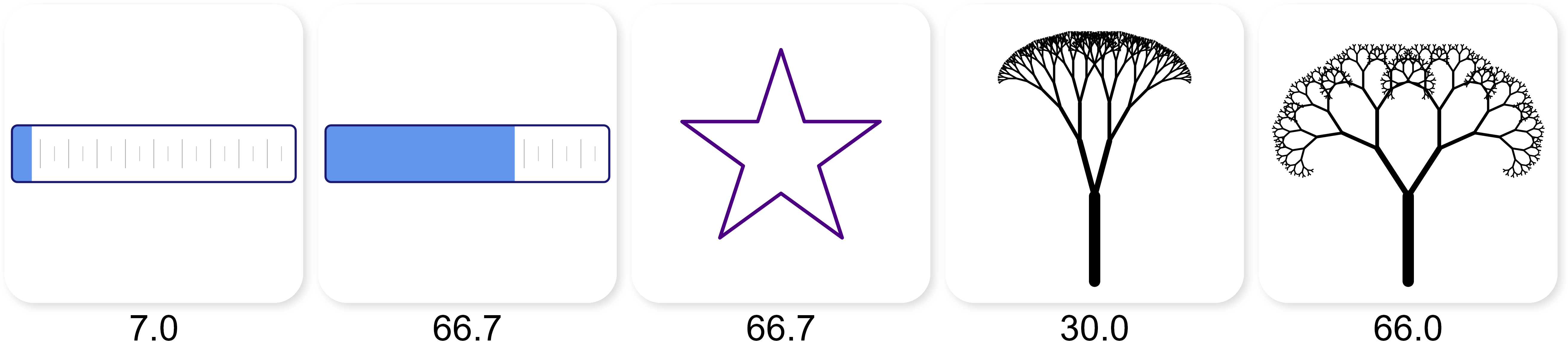}
    \caption{Some glyphs allow a very accurate reading of the absolute value of the parameter $x$, some only approximately, and for some, the absolute value is very difficult, if not impossible, to estimate.}
    \label{fig:absolute-relative}
\end{figure}
A progress bar (\cref{fig:absolute-relative}, first and second glyphs), which fits within the one-inch dimension, is a special case of a malleable glyph.  With a progress bar, it is possible to estimate very well the absolute value of its completeness, similar to a speedometer-style indicator or the one-handed clock in \cref{fig:one-handed-clock}.  For other glyphs, estimating the absolute value is more difficult, and it will certainly be possible to construct glyphs that intentionally provide few clues for determining the absolute value of $x$ (such as the fractal tree in \cref{fig:absolute-relative}, fourth and fifth glyph), but still allow for high glyph resolution and reliable relative comparisons between two glyphs.

The ability of a glyph to communicate the absolute value of its parameter may not always be an advantage.  Accurately reading the value of progress tends to be a source of frustration for users, for example, when executing large tasks (e.g. operating system upgrade) where it is difficult to estimate the duration of parts of the task and progress ``stalls'' observably at some points.  In computer games, the health of opponents is often displayed using a progress bar, but some games opt for less readable ways of communicating the wear of an opponent, where the relative change in damage is visible, and the player has a rough idea of the opponent's health, but an accurate, numerical reading is not desirable for the player experience.  Insights from studying malleable glyphs can be useful in such, and similar applications.

We leave for future work the definition of a precise way to evaluate and quantify this ability of a particular glyph to communicate in \textbf{absolute vs. relative} terms, i.e., to what extent a glyph is capable of unambiguous mapping to the interval $x\in[0.0, 100.0]$ at the same resolution $\mathcal{A}(G)$ or $R(G)$.


\begin{figure}
    \centering
    \includegraphics[width=\linewidth]{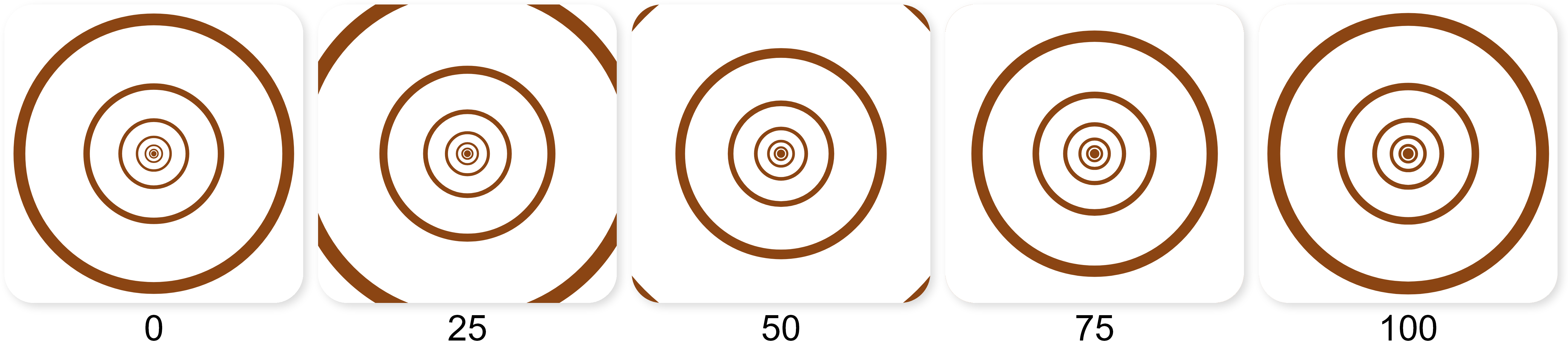}
    \caption{Simple glyph with the Shepard property: the glyph appearance is identical for $x=0.0$ and $x=100.0$, and the glyph is therefore ``circular''.}
    \label{fig:shepard-circle}
\end{figure}

A Shepard tone \cite{Shepard1964} is a musical tone composed of multiple appropriately placed fundamental frequencies so that it is possible to increase (or decrease) its pitch, and the tone returns to the pitch it was previously at. The increase (or decrease) can be applied periodically and continuously to achieve what is known as a Shepard-Risset glissando -- the listener perceives the sound to increase steadily and without end.  A corresponding property is present in the \textbf{Shepard malleable glyph} in \cref{fig:shepard-circle}.  The appearance of the glyph for parameter $x=0.0$ and for $x=100.0$ is completely identical, and the glyph transitions smoothly between these values so that no beginning or end of parameter increase can be determined.

It is noteworthy that many controls in today's world acquire this \emph{Shepard property}.  This is the case, for example, for physical audio volume controls in consumer electronics or in-car infotainment devices, which used to have a zero position and a maximum position (typically labeled 10), but now have no end and no beginning.  Similar controls are awaited by social network feeds or perhaps product catalogs on e-stores. Currently, similar applications use a vertical scroll bar that periodically (and somewhat randomly and surprisingly) changes its extent.  Research of and experimenting with malleable glyphs (those with the Shepard property) may provide inspiration for new UI elements that better correspond to the ``infinity'' of many current UIs and feeds.


We propose to evaluate glyphs on a computer screen as described in \cref{sec:glyph-evaluation}.  However, the original purpose of the glyphs is to be printed on cards for user testing.  The cards (sketched in \cref{fig:q-sorting-motivation}) are rectangular in shape, so the glyphs that should be usable for this purpose must be invariant with respect to a rotation of $180^\circ$. Other purposes may require the cards to be square, so the glyphs should be invariant to rotation by multiples of $90^\circ$, and still other applications may want glyphs invariant to rotation by arbitrary angles.  For example, the three-handed clock in \cref{fig:three-handed-clock}, which outperformed greatly the capabilities of all the other glyphs presented due to its use of (unallowed) digits and orders, is not invariant to rotation at all, and even if the orientation of the dial is visually indicated, comparing the glyphs after rotation would be difficult and error-prone.  Thus, \textbf{invariance to rotation} is clearly another property that will be meaningful to consider and evaluate for the proposed malleable glyphs in the future.


The three views on glyphs proposed in this chapter are, so far, only a rough sketch of how to classify glyphs and how to evaluate them on various additional scales.  As new glyph designs and their combinations are observed, it will be appropriate to propose and explore new perspectives on glyph taxonomy and evaluation.  An example of one promising view that we want to explore, once we have a sufficient collection of diverse enough glyphs, is whether the observers see them as natural metaphors of `greater' or `lesser' in the sense proposed by Lakoff and Johnson \cite{Lakoff1980} or whether other metaphors are more suitable.

\section{The Malleable Glyph Challenge}

Our intention in writing this text is to offer fellow colleagues a \textbf{new sport} -- with strict and deliberately restrictive rules (as is common in sports), with the opportunity for home work, self-preparation, and self-evaluation, and with the possibility of public and fair comparison with others.

\subsection{Python Tool for Creating Malleable Glyphs}

A straightforward way for anyone to create a new malleable glyph design is to use the mglyph Python library we have published.  The library unifies the way glyphs are rendered, supports both vector and raster rendering, and is primarily intended for prototyping glyphs in a Jupyter Notebook environment, including Google Colab.  A simple example of using the library is here:
\begin{python}
import mglyph as mg
def horizontal_line(x:float, canvas:mg.Canvas):
    canvas.line((mg.lerp(x, 0.01, -1), 0), 
                (mg.lerp(x, 0.01, 1), 0),
                color='mediumblue', width='30p')
mg.show(horizontal_line)
\end{python}
The example defines a glyph called \pythoninline{horizontal\_line}, which takes the form of a blue horizontal line that has a minimum length for $x = 0.0$ and is as long as the whole glyph area for $x = 100.0$. Working with the library is explained from the basics to more advanced concepts in the tutorial, which can be found on \href{https://mglyph.net}{our website} and also works in the colab environment, so it does not necessarily require installing any software on one's computer.

\subsection{Exchange Format}

One way to share designed glyphs between glyph designers and scientists investigating their properties would be to share their source code based on the described library.  However, we want to make it possible to create glyphs using other tools and approaches and to ensure complete certainty about how the vector description of the glyph is rasterized. To exchange glyphs for mutual comparison and further exploration, we propose to rasterize them into images with \emph{high enough} resolution. 

At the present time, and when considering a test scenario using a conventional smartphone, tablet, or desktop display in the usual position of the observer's head relative to the device, we consider the resolution of 500 pixels per inch (PPI) to be sufficiently high.  This is a value that exceeds the resolution of even the high-end consumer electronics displays, and it exceeds the retinal resolution of human eyes for normal viewing distances \cite{Ashraf2024,Campbell1966}.  We assume that reducing the resolution through an appropriate filter to the target display resolution will not degrade the perceived image quality.

The simple glyphs that we have experimented with so far (and which are mostly shown as examples in this article) achieve a resolution $R(G)$ of up to around 800.  Thus, for basic experimentation, it may be appropriate to export about 2\,000 uniformly sampled glyph sizes; for a truly reliable evaluation, it may currently be sufficient to export 10\,000 uniformly sampled sizes.  Non-uniform sampling may also be considered to save memory, but schemes for such a procedure will be the subject of future work.

For ease of manipulation of the exported glyph, we propose to compress the exported images stored in the lossless PNG format into a single ZIP archive along with a single \texttt{info.json} file containing the metadata of the glyph and a list of all the extracted images and their corresponding $x$ values:

\begin{lstlisting}[language=json]
{ "name": "Horiz. Line", 
  "short-name": "line", 
  "author": "Jane Designoff", 
  "e-mail": "jd@mglyph.net", 
  "version": "1.5.3", 
  "creation-time": "2025-03-07 09:31:46.397244", 
  "images": [ [ 0.00, "00000.png" ], 
              [ 0.01, "00001.png" ], 
              [ 0.02, "00002.png" ], ...   ...
              [ 100.00, "10000.png" ]] }
\end{lstlisting}
The \texttt{mglyph} library mentioned above supports this export, and, for example, this code exports 10\,001 uniformly sampled sizes of a given glyph:
\begin{python}
mg.export(horizontal_line, 
          xvalues=np.linspace(0.0, 100.0, 10001),
          name="Horiz. Line", short_name='line',
          path="glyphs/Horizontal Line.mglyph",
          author="Jane Designoff", 
          email="jd@mglyph.net", version="1.5.3")
\end{python}

\subsection{Online Self-Evaluation Tool}

For self-evaluation of glyphs on the part of the glyph designer or other interested parties, we provide a tool that allows the evaluation of one or more glyphs in a single session to ensure cross-comparability.  When using our self-evaluation tool, the glyphs are not uploaded to any server, but the glyph data is stored in the user's web browser.  On the one hand, this means that the evaluation and comparison are not subject to any central control and do not guarantee the inter-comparability of the measurements; on the other hand, it means ``privacy'' when developing new glyph designs and exploring their properties.  In the metaphor of \emph{newly defined sport}, the use of a self-evaluation tool corresponds to home training with one's own time measurement or other means of evaluating athletic performance: as preparation for public competition and as a pre-qualification tool, this method of evaluation seems to be appropriate and useful, although it is surely not credible enough to establish a new world record.

The self-evaluation tool mentioned is available from the malleable glyph homepage \href{https://mglyph.net}{\texttt{mglyph.net}}.

\subsection{The Challenge}

\emph{If you are a graphic designer, a researcher in computer graphics, visualization, visual cognition, or computer vision, or anyone interested in the problem of designing the best-performing glyph, please join us in an effort to combine the tiny streams of diverse professional backgrounds and personal experiences into a stronger stream.  Take inspiration from the malleable glyphs that are currently available and suggest combinations of them, innovations of them, or completely different ways to conceive a malleable glyph.}

\emph{As organizers of the challenge, we commit to evaluate promising glyphs carefully and impartially, and at appropriate times to work with the authors of innovative, high-performance, aesthetically appealing, and otherwise beneficial glyphs to describe, categorize, and analyze them, and jointly publish the results to the scientific and design community.  Together with other interested parties, we want to work on refining the definition of a malleable glyph and the rules of glyph eligibility, and on the eventual announcement of additional useful categories of malleable glyphs.}

\begin{tcolorbox}
    \emph{In the initial version of the challenge, we define the malleable glyph as a static graphical design that fits into a square area of size $1\times1\,\mathrm{inch}$ with corners rounded by a radius of 10\,\% of the glyph width, that is parameterized by $x\in[0.0, 100. 0]$ so that a human (or even artificial) observer can distinguish in a pair of glyphs which one has the higher and lower parameter $x$ or whether they are equal in this respect, and the glyph preserves the illiteracy rule described above so that the question the observer asks is ``how much?'' rather than ``how many?''}
\end{tcolorbox}

\section{Conclusions and Future Work}

The Malleable Glyph is intended to serve graphic designers and visualization researchers to exchange inspiration and share their experiences.  The goal was to define clear rules defining what exactly a malleable glyph is but to be as permissive as possible at the same time.  The goal is also to minimize `boilerplate activities' so that someone with an idea of how to graphically communicate scalar value can implement and evaluate their idea with minimal effort, focusing just on their idea and its graphical appearance.  In the near future, we intend to respond to requests and comments from glyph designers and to further develop the tools for glyph creation and evaluation.  We will also closely observe the implementation of the defined rules and, if necessary, refine the rules to make them unambiguous, but at the same time still as permissive as possible.

From the glyph set, which is steadily growing, we hope to explore the properties and capabilities of neural networks used in computer vision.  Malleable glyphs were created to enable visual communication with a human subject.  We are extremely curious to see what similarities and differences machine learning models will exhibit in the perception of partial elements of glyphs compared to human subjects.  We are curious to what extent neural networks will be able to truly `understand' the concept of visual digits and orders (which we have defined as ineligible for glyphs, but are still possible to experiment with).  We are curious about the ability of machine learning models to generalize and learn to discriminate a new glyph based on knowledge learned from a multitude of other glyphs.  We are curious to what extent machine learning models will be sensitive to the Weber-Fechner logarithmic law.  We believe that the concept of a malleable glyph can lead to interesting insights into the properties, behavior, and potential of neural networks in computer vision.



\bibliographystyle{IEEEtran}
\bibliography{the_bibliography.bib}

%






\end{document}